\documentclass[runningheads]{llncs}
\usepackage{graphicx}
\usepackage{times}

\begin{document}

\title{Genome-Wide Epigenetic Modifications as a Shared Memory Consensus Problem}
%
\author{Sabrina Rashid\inst{1} \and Gadi Taubenfeld\inst{2} \and Ziv Bar-Joseph\inst{1}}
%
%
\institute{
Carnegie Mellon University, Pittsburgh, USA
\and
The Interdisciplinary Center, Herzliya, Israel
}
\maketitle              


\begin{abstract}
A distributed computing system is a collection of processors that
communicate either by reading and writing from a shared memory or
by sending messages over some communication network.
Most prior biologically inspired distributed computing algorithms rely on message passing as the communication model. Here we show that in the process of genome-wide epigenetic modifications cells utilize their DNA as a shared memory system. We formulate a particular consensus problem, called
\emph{the epigenetic consensus problem},
that cells attempt to solve using this shared memory model, and then present algorithms, derive expected run time and discuss, analyze and simulate improved methods for solving this problem. Analysis of real biological data indicates that the computational methods indeed reflect aspects of the biological process for genome-wide epigenetic modifications.
\end{abstract}

\section{Introduction}

Like virtually all large-scale computing platforms, cellular and molecular systems are mostly distributed consisting of entities that interact, coordinate, and reach decisions without central control \cite{navlakha2015distributed}.
To date, coordination in such processes was almost always discussed in the context of message passing. Famous examples include neural networks \cite{bishop1995neural} where neurons communicate by passing messages along synapses, cellular decision processes where cells communicate by secreting proteins \cite{afek2011biological,singh2016distributed} and protein interaction networks where proteins physically interact to achieve a common goal \cite{weigt2009identification}.

While message passing is indeed a useful and dominant method for distributed biological computing, some biological processes utilize another method for coordination. This method is similar to the method of communicating via shared memory, studied intensively in distributed computing, and utilizes modification to the DNA which can be sensed (read) by other participating entities. This process is termed \emph{epigenetics} and involves several different groups of proteins which could largely be divided into three groups: \emph{readers, writers, and erasers} \cite{peterson2004histones}.

Epigenetic refers in part to post translational modifications of the histone proteins on which the DNA is wrapped. Such modifications play an important role in the regulation of gene expression and so are themselves highly regulated and consistent across large stretches of the genome \cite{gunther2010epigenetic,peterson2004histones}.  In particular, when switching between cell states (for example, when facing stress or during development), a coordinated set of modifications are required such that expression programs required for these states can be executed.

Stretches of about 170 base pairs are wrapped around octets of histone proteins to form nucleosomes \cite{diesinger2009depletion}. These and other DNA-associated proteins make up the \emph{chromatin}.
Researchers have been cataloging chromatin proteins and their modifications,
in an effort to segregate chromatin's complexity into discrete numbers of chromatin states \cite{berger2007complex}.
Chromatin-state mapping promises to reveal many secrets of genome function,
how cells inherit acquired states, how chromatin directs functions such as transcription and RNA processing,
and how chromatin alterations contribute to disease response and progression.

There are several types of histone modifiers that regulate transcriptional activity by changing chromatin states. As mentioned, these modifiers can be broadly categorized into three classes: readers, writers, and erasers \cite{peterson2004histones}. As their names suggest, writers can add specific modifications (but cannot erase existing ones), while erasers can remove specific modification but cannot add others. Readers are the executers, and while their activity is the goal of these modifications we would not discuss them from now on since their involvement in the process is limited to the outcome whereas we are focused here on understanding the process of reaching consensus as discussed below.

Several specific types of histone modifiers have been discovered including acetylation, methylation, phosphorylation, ubiquitination, and so on \cite{peterson2004histones}. While each has its own interpretation, in this paper we generalize all of them to focus on the interplay between writers and erasers rather than on the specific histone modification type. However we note that a number of modifiers (including both erasers and writers) were shown to act on the exact same locations \cite{wang2008combinatorial}, which corresponds to a number of processors that may access the same memory location in a shared memory system Coordination between writers and erasers for these sites is required to maintain a consistent state across the genome.

Although a lot of recent work in the biological literature has focused on histone modifications and the specific proteins that regulate this process, to the best of our knowledge no prior work discussed the issue of global coordination between writers and erasers. We often see \emph{exactly the same} histone marks for a stretch of DNA  \cite{gunther2010epigenetic}. However, how such consensus is reached for a particular mark is still not clear. Our goal in this paper is to understand better how such consensus may be reached.

To this end, we model the histone modifiers as two different types of writer processors and two different types of eraser processors that communicate by accessing a shared memory array (a stretch of DNA), and for such a setting formally define the \emph{epigenetic consensus problem}. We first discuss a simple algorithm for solving the problem and then present a more sophisticated algorithm, that better matches various biological assumptions, and discuss its run time both theoretically and in simulations.

\section{Model}
\label{sec:model}
We assume a shared memory array with $N$ cells, which corresponds to
a linear DNA stretch with $N$ total histones that can be modified
($N$ is a large number -- and these histones cover hundreds of thousands of DNA bases).
Recent biological results indicate that such stretches of DNA, often anchored by CTCF bindings sites, are likely to be jointly modified when switching between cell states \cite{tang2015ctcf}. Thus, our focus here is on achieving a consensus assignment for such stretches when cells need to switch between different biological states.


Since our processors mimic memory-less proteins, we further assume that each processor has only two bits of memory which, for example, prevents it from counting. While there could be several types of modifications, many of them can co-exist (each changing a different histone residue). For a specific residue most of the modification are restricted to two possible values and so we assume here that only two values can be written to each location, denoted by 0 and 1. Again, following the biological model we allow each processor to be either a writer or an eraser \cite{goldberg2007epigenetics}.

We assume that two types of writers, 0-writers and 1-writers, are assigned to each DNA stretch. $W_0$ denotes the number of 0-writers and $W_1$ denotes the number of 1-writers.
Similarly, we have sets of erasers for 0 and for 1, such that 0-erasers can only erase 0 and 1-erasers can only erase 1. Following recent studies, \cite{torres2015functional} we assume that the generation of writers and erasers is similarly regulated and so number of 0-erasers, $E_0 = W_1$ and number of 1-erasers, $E_1 =W_0$.

To switch between states, cells transcribe (generate) new writers and erasers for the modification needed (for example, when changing from 0 to 1). When changing from 0 to 1, the new 1-writers will usually outnumber the existing 0-writers. However, we cannot expect these 0-writers to completely disappear, at least not within the time scales needed for changing the state. Thus, we assume that at any given time the number of one type of writers is larger than the number of the other type, and the convergence time of our algorithms will depend on this assumption.

Each of the $N$ locations in the shared memory can be in one of three states: Empty (V), 0 or 1. A state transition can occur from V to 0 (1) by a 0 (1) writer and from 0 (1) to V by a 0 (1) eraser. However, a transition from 0 (1) to 1 (0) cannot occur. That is, a 0 state needs to be erased first and only then can be written by a 1-writer.
It is assumed that reading a cell and then possibly updating it is done as one atomic step.
In addition, it is assumed that all locations are initially empty. This assumption is based on studies showing that all marks are completely erased before being rewritten in certain reprogramming events \cite{Meyenn2015Era}.
To address cases where there is no global erasing of markers, in Section~\ref{sec:Discussion}, we relax this assumption and explain how it can be easily removed. More precisely, we show in Section~\ref{sec:Discussion} that with a tiny change to our algorithm
it satisfies the self-stabilization property, that is,
starting from any initial assignment of values to the memory locations (i.e., starting from any configuration)
the algorithm always produce the desired final result \cite{Dij74,Dolev2000}.

The shared memory locations are \emph{anonymous}. That is,
they do not have names that the processors a priori agreed upon \cite{Tau2017podc}.
Each one of the processors (i.e., writers and erasers) starts by accessing a random location in the array.
Thereafter, in its next step a processor may access one of the two locations which are
adjacent to the location it has accessed in its previous step.
One may view the processors as mobile agents that are moving between locations in the shared array.%
\footnote{In particular, a processor does not have to
remember the address (name) of the last memory location it has accessed which will require $log~N$ bits of local memory.
It only needs to remember one bit which represents the direction in which it is moving.}

The writers and erasers do not have a (global) sense of direction.
That is, they do not a priori agree on which side of the array
is the left side and which side is the right side.
We assume that writers and erasers are \emph{asynchronous},
nothing is assumed about their relative speed.
The time efficiency of an asynchronous algorithm is
often of crucial importance. However, it is difficult
to find the appropriate definition of time complexity, for
systems where nothing is assumed about the speed of the processors. Thus, for measuring time, we will assume that a single step of a processor takes at most one time unit.

Assume that a computation is taking place through time
and that every step of every processor takes some amount in the interval (0,1]. That is, there is an upper bound 1 for step time but no lower bound. Thus, for example, if during some period of time, where two processors are taking steps, one processor takes 100 steps while the other takes 5 steps, then the total time that has elapsed is at most 5 time units.

Under the assumption that a single step of a processor takes at most one time unit,
the time complexity of an algorithm is defined as the maximum number of time units
(also called ``big steps'') that are required for the algorithm to converge (i.e., to terminate)
\cite{CZ89,HSbook2008,RaynalBook2013}.
For the rest of the paper, by
a step of the algorithm, we mean a ``big step'' which takes one time unit
and where each correct processor that started participating in the algorithm has taken
at least one step.
For example, in Section 4
we derive the expected number of big steps (i.e., time units)
for our consensus algorithm to converge.

\section{The epigenetic consensus problem}
The {\em epigenetic consensus problem\/} is to design an algorithm in which all
processors reach agreement based on their initial opinions.
In our context, reaching agreement is expressed by
guaranteeing a consensus outcome of either 0 or 1 for all $N$ memory locations.
An epigenetic consensus algorithm is an algorithm that produces such an agreement, assuming only writers and erasers as defined previously.

More formally the problem is defined as follows.
There are a fixed number of $W_0$ 0-writers,
$W_1$ 1-writers,
$E_0$ 0-erasers, and
$E_1$ 1-erasers.
Recall that $W_1 = E_0$ and $W_0 = E_1$.
Initially, each one of the $N$ memory locations is empty, and upon termination,
the value of each location is either 0 or 1.
The requirements of the epigenetic consensus problem are that there
exist a {\em decision value}
$v\in \{0,1\}$ such that,
\begin{itemize}
\item
    \emph{Agreement}:
    With probability 1,
    the value of each one of the $N$ memory locations is eventually $v$, and
    does not change thereafter.
\item
    \emph{Majority}: When there is a strong majority of $v$-writers then,
    with high probability, the final decision value (i.e., the final value of each one of the locations)
    is $v$.
\item
    \emph{Validity}: The final value of each location is a value of some writer.
\end{itemize}
We point out that the first requirement has two
parts. The first (the agreement part) is that all $N$ memory locations eventually contain the same
value, and the second (the termination part) is that eventually the memory locations
do not change their agreed upon value.
In the second requirement, a strong majority of $v$-writers means that
$W_v / W_{1-v} \geq 3$ (recall that $v\in\{0,1\}$).
We use a ratio of 3 to 1 here. However, we note that while recent studies have shown that over-expression of specific writers leads to change of consensus value, as we assume, the actual ratio in real biological processes has not yet been fully determined \cite{Cano2016ratio}.
The third requirement ensures that if $W_0 = 0$ then
the decision cannot be on the value 0, and similarly if
$W_1 = 0$ then the decision cannot be on the value 1.
Thus, it precludes a solution which always decides 1 (resp. 0).
The consensus problem defined above is also called \emph{binary} consensus
as the decision value $v$ is either 0 or 1.
A generalization of the
problem where the $v$ is taken from a larger set is not considered in this paper.

The consensus problem is a fundamental coordination problem and is at the core of many algorithms for distributed applications. The problem was formally presented in \cite{PSL80,LSP82}. Many (deterministic and randomized) consensus algorithms have been proposed for shared memory systems. Few examples are
\cite{Abrahamson1988,ASS2002,aspnes1990fast,FMT93,LA87,Plo89,SSW91}.
Dozens of papers have been published on solving the consensus problem in various
messages passing models. Few examples are \cite{DDS87,DLS88,Fi83,FLM86,FLP85}.
For a survey on asynchronous randomized consensus algorithms see \cite{Aspnes2003}.

\section{Algorithms}
\subsection{A naive algorithm}
Before we present our main algorithm,
we first discuss a straightforward but non-desirable solution which is easy to analyze.
In this solution, the erasers do not participate. Writers compete on writing the leftmost cell (i.e, memory location)
and the value
written into that cell becomes the final agreed upon value. This is done as follows.
Assume $v$ is the written value. The $v$-writers continue writing $v$ into all the
cells. The major downside of this solution is  that the probability of ending with the majority value is
$p = W_1/(W_0+W_1)$
(assuming a 1 majority) which is usually very dangerous for cells since there is a constant probability of not reaching the desired state.
Furthermore, this solution assumes that all the writers have the same
orientation (i.e., they a priori agree on which side is the left side) which is an unacceptable assumption. We present below a much better solution.

\subsection{The epigenetic consensus algorithm}
While the algorithm above will finish in $\Theta(N)$ steps, as mentioned it may not lead to the desired outcome. Instead, we propose to rely on recent biological observations that indicate that stretches of consecutive 1's (resp.\ 0's) are locally extended until they reach other stretches of 1's
(resp.\ 0's) \cite{Becker2013move}. Based on this we propose the following algorithm. Let $v\in\{0,1\}$.

\begin{itemize}
\item
Each of the writers and erasers starts at a random location.
Their direction is also chosen randomly.
\item
\emph{Rule for a $v$-writer}:
The writer starts moving in the chosen direction.
If it sees an empty cell, it writes $v$ and moves on to the next cell.
When a $v$-writer reaches the end of the stretch, it reverses its direction and continues.
\item
\emph{Rule for a $v$-erasers}:
The $v$-eraser starts moving in the chosen direction.
When it sees a value $v$ which is preceded by the value $1-v$,
it erases the $v$.
Otherwise, it just moves on.
When a $v$-eraser reaches the end of the stretch, it reverses its direction and continues.
\end{itemize}
When the values of two consecutive non-empty cells are different, we call that a \emph{collision}.
The algorithm will run until all the cells' values are non-empty and until all collisions are resolved.
Intuitively, each time there is an empty cell after a collision is resolved, assuming $W_1 > W_0$,
the probability that the value 1 will be written is higher,
and thus the algorithm will eventually converge.
Another version of the algorithm we considered, forces a writer to spin (wait)
when it notices a collision.
Simulations indicate that the spinning version is more efficient than the non-spinning one,
but this version is harder to analyze.

\section{Analysis: Preliminaries}
\label{sec:AnalysisPreliminaries}
We prove that the algorithm satisfies the \emph{majority} requirement,
and compute the expected runtime.
We do that by applying known results about biased random walks in one dimension.
More precisely, we use the following well-known solution to the
\emph{Gambler's ruin problem} \cite{Feller1959,Ross1983}.

\subsection{The gambler's ruin problem with ties allowed}
\label{subsecGRP}
Consider a gambler who at each play of the game has probability $p$ of winning one unit,
probability $q$ of losing one unit, and probability $r$ of not winning or losing ($p+q+r=1$; $0<p,q<1$).
(In gambling terminology, when $r>0$ a bet may result in a tie.)
Assume successive plays of the game are independent, what is the probability that starting with $0\leq i\leq N$ units,
the gambler's fortune will reach $N$ before reaching 0.

\begin{lemma}[The gambler's ruin lemma]
\label{lemma:GamblerRuinLemma}
Let $f_i$ denotes the probability that starting with $i$ units, $0\leq i\leq N$,
the gambler's fortune will eventually reach $N$. Then, assuming $p\neq q$,
\[ f_i =  \frac{1-(q/p)^i}{1-(q/p)^N} \]
\end{lemma}
Lemma \ref{lemma:GamblerRuinLemma} is true regardless of the value of $r$
(i.e., regardless of whether ties are allowed or not).
As $N\rightarrow\infty$,
if $p> q$, there is a positive probability that the gambler's fortune will converge to infinity;
whereas if $p< q$, then, with probability 1, the gambler will eventually go broke when playing against
an infinitely rich adversary.
(When $p=q$ and $r=0$, $f_i = i/N$.)
Even though casino gamblers are destined to lose, some of them enjoy the process.
Lets figure out how long their game is expected to last.

\begin{lemma}[Expected playing time]
\label{lemma:Expected Playing Time}
Let $E_i$ be the expected number of bets before going home (broke or a winner),
starting with $i$ units, $0\leq i\leq N$. Then, assuming $p\neq q$,
\[
E_i = \left( \left( \frac{N}{p-q}\right) \left[ \frac{1-(q/p)^i}{1-(q/p)^N} \right] - \frac{i}{p-q} \right) \left( \frac{1}{p+q} \right)
\]
\end{lemma}
Without ties (i.e., when $r=0$) the right term equals 1.
The expression is much simpler in the following cases.
When $p>1/2$, $r=0$ and both $i$ and $N$ are large, $E_i \sim {(N-i)}/(2p-1)$.
This seems to make sense since the gambler is expected to \emph{win} $1p - 1(1-p) = 2p-1$ units on
every bet and starting with $i$ units, the gambler needs to win additional $N-i$ units.
On the other hand, when $p< 1/2$, $r=0$ and $N-i$ is large, $E_i \sim i/(1-2p)$. This seems
to make sense since the gambler is expected to \emph{lose} $1 (1-p) + (-1)p = 1-2p$ units on
every bet and the gambler started with $i$ units.
(When $p=q$ and $r=0$, $E_i = i(N-i)$.)

Finally, the sum of the probabilities that, starting with $i$ units,
the gambler's fortune will reach $N$ or the gambler will eventually go broke
is known to be 1, so we need not consider the possibility of an unending game.
That is, for every $p>0$, the probability that the game never ends is 0.
\subsection{Chernoff Bound}
One can encounter many flavors of Chernoff bounds.
We will use the following version,
\begin{theorem}[Chernoff Bound~\cite{Chernoff1983}]
\label{thm:ChernoffBound}
Let $X = \sum_{i=1}^n X_i$
where $X_i = 1$ with probability $p_i$ and $X_i = 0$ with
probability $1-p_i$, and all $X_i$ are independent. Let $\mu = E(X) = \sum_{i=1}^n p_i$.
Then,
$$
Pr(X \geq (1 + \delta)\mu) \leq e^{-\frac{\delta^2}{2+\delta}\mu}
~~~for~ all~ \delta > 0
$$
\end{theorem}

\subsection{Additional assumptions about the model}
We state below a few assumptions that capture important aspects of our model,
and simplify the analysis of the epigenetic consensus algorithm.
%
Recall that it is assumed that $W_1 = E_0$ and $W_0 = E_1$.
The value that is written into an empty memory location depends on the
ratio between the 1-writer and 0-writers at that location.
In an asynchronous system it is not possible to exactly tell what this ratio is at
any given time. However, this ratio is, of course, going to be effected by the
overall ratio between the different types of writers.
Thus, to abstract away from all the physical details (such as, the time it takes to write, erase, move to the next location, etc.),
which may affect the current location of a process, we assume the following,
\begin{quote}
\emph{
Assume that a value (0 or 1) is written at a certain location.
The probability that the value written in that location is $v\in\{0,1\}$ is $W_v/(W_0 + W_1)$.
}
\end{quote}
The assumption implies that a value that is written
does not depend on past history. Thus, there are no probabilistic dependencies between the values
written into different (empty) memory locations.
Let $X_v$ be the number of values $v\in\{0,1\}$ that are written into the $N$ memory locations for the first time
(i.e., for each location we consider the first value written into it).
By definition, $X_v + X_{1-v} = N$. By the assumption above, $E(X_v) = (N \times W_v) /(W_0 + W_1)$.

Given physical space constraints, when resolving a collision on the chromatin,
only one value in one of the two adjacent locations participating in the collision can be erased,
but not both values. That is, because of their physical size, either a 0-eraser or a 1-eraser may
observe a specific collision but not both at the same time.
Given this biological observation, from now on we will assume the following,
\begin{quote}
\emph{
An eraser reads two adjacent locations, and possibly erases one of them,
in one atomic step.
Thus, when a collision is resolved,
only one value in one of the two adjacent memory locations
participating in the collision can be erased.
Furthermore, in case of a collision,
the probability that the value erased is $v\in\{0,1\}$ is $E_v/(E_0 + E_1)$.
}
\end{quote}
Thus, there are no dependencies between erasers which attempt
to access overlapping (or the same) collisions concurrently.
Finally, for simplicity, we assume that before the first
value is erased, each one of the $N$ locations is written at least once.


\section{Analysis: The probability of reaching agreement on the majority value}
We prove that the epigenetic algorithm satisfies the \emph{majority} requirement.
%
\begin{theorem}[Satisfying the majority requirement]
\label{thm:majority}
Assume $W_1 > W_0$. The probability that the final decision value is 1 is more than
$$
\left(1-(W_0/W_1)^{4 W_0 N/(W_0 + W_1)}\right) \times \left(1- e^{-W_0 N/(3(W_0 +W_1))}\right)
$$
\end{theorem}
\begin{corollary}
\label{corollary:majority}
Assume $W_1 / W_0 \geq 3$. The probability that the final decision value is 1 is more than
$$
\left(1-(1/3)^{N}\right) \times \left(1- e^{-N/12}\right)
$$
\end{corollary}
Thus, when there is a strong majority of $v$-writers then, with high probability,
the final decision value is $v$.
We prove the theorem by applying the known result about the \emph{Gambler's ruin problem} as
captured in Lemma~\ref{lemma:GamblerRuinLemma}, and by using Chernoff bound (i.e., Theorem~\ref{thm:ChernoffBound}).

For the rest of the section we prove Theorem~\ref{thm:majority}.
Let us focus on \emph{update-steps} in which, in an attempt to resolve a collision,
a value is erased and then a value is written in the same location.
An update-step may result a change in the number of 1's (and hence also of 0's).
There are three such types of update-steps which we will name \emph{win}, \emph{lose} and \emph{tie}.
A win step is when a collision (of 01 or 10) is changed into 11.
A lose step is when a collision (of 01 or 10) is changed into 00.
A tie step is when a collision is not changed (a value is erased and then the same value is written).
So, the number of 1's increases by one in a win step, it decreases by one in a lose step and
it does not change in a tie step.

Recall that $W_1 = E_0$ and $W_0 = E_1$, and the probability of erasing 0 or writing 1 is
$W_1/(W_0 + W_1)$.%
Let $P$ be the probability that the number of 1's increases (and 0's decreases) in an update-step;
let $Q$ be the probability that the number of 0's increases (and 1's decreases);
and let $R$ be the probability that the number of 1's and 0's does not change.
Then,
$$
P = \left(\frac{W_1}{W_0 + W_1}\right)^2~~~;~~~Q= \left(\frac{W_0}{W_0 + W_1}\right)^2~~~;~~~R=1-P-Q.
$$
%
%
Assume that all the memory locations are not empty. Let $i$
denotes the initial number of 1's, after each one of the $N$ locations is written once.
Notice that, by definition, the initial number of 0's is $N-i$.
The question that we are interested in
is, what is the probability that starting with $i$ values of 1 (after each of the locations is written once),
the final decision value (i.e., the final value of each one of the $N$ locations) is 1?

This question is identical to the question asked in the gambler's ruin problem that we just analyzed.
Going broke is analog to reaching agreement on 0, where going home winner with a fortune of
$N$ is analog to reaching agreement on 1.
Winning one unit with probability $p$ is analog to a win step with probability $P$,
losing one unit with probability $q$ is analog to a lose step with probability $Q$, and
not losing or winning with probability $r$ is analog to a tie step with probability $R$.
Finally, the gambler starting with $i$ units is analog to assuming that the initial number of 1's is $i$.
Thus, by Lemma~\ref{lemma:GamblerRuinLemma} we get,
\begin{lemma}
\label{lemma:write-erase}
Let $f_i$ denote the probability that starting with initially
$i$ values of $1$, $0\leq i\leq N$,
the final decision value (i.e., the final value of each one of the $N$ locations) is $1$.
Then, assuming $P\neq Q$,
\[ f_i =  \frac{1-(Q/P)^i}{1-(Q/P)^N} = \frac{1-(W_0/W_1)^{2i}}{1-(W_0/W_1)^{2N}} \]
\end{lemma}
%
%
For the rest of the section, let $\mu_0$ denotes the expected initial number of 0's.
Clearly, $\mu_0 = W_0 N/(W_0 + W_1)$. Thus, when assuming that $W_1 > W_0$, we get that  $0\leq 2\mu_0\leq N$.
By Lemma~\ref{lemma:write-erase},
\begin{lemma}
\label{lemma:randomWalk}
Assume $W_1 > W_0$.
The probability that the final decision value is the majority value 1,
when starting with at least $2 \mu_0$ values of 1, is at least
$$
\frac{1-(W_0/W_1)^{4 \mu_0}}{1-(W_0/W_1)^{2N}} \geq 1-(W_0/W_1)^{4 \mu_0}
$$
\end{lemma}

\subsection*{The probability of starting with at least $2 \mu_0$ values of 1}
In Lemma~\ref{lemma:randomWalk}, we computed the probability that the final decision value is the majority value $1$,
conditioned on starting with at least $2 \mu_0$ values of $1$.
Using Chernoff Bound (Theorem~\ref{thm:ChernoffBound}), it is possible to
compute the probability of starting with at least $2 \mu_0$ values of 1.
\begin{lemma}
\label{lemma:usingChernoff}
Assume $W_1 > W_0$.
Let $X_1$ be the initial number of 1's. Then,
$$
Pr\left(X_1 \geq 2 \mu_0 \right) > 1- e^{-\mu_0 /3}
$$
\end{lemma}
%
%
\proof
Recall that $\mu_0$ is the expected initial number of 0's ($\mu_0 = W_0 N/(W_0 +W_1)$).
Let $X_0$ be the initial number of 0's.
By substituting $\mu_0$ for $\mu$ and 1 for $\delta$ in Theorem~\ref{thm:ChernoffBound}
(Chernoff Bound) we get that,
\begin{equation}\label{B}
Pr\left(X_0 \geq 2 \mu_0\right) \leq e^{-\mu_0/3}
\end{equation}
That is, the probability of starting with at least $2 \mu_0$ values of 0 is at most $e^{-\mu_0/3}$, which implies that,
\begin{equation}\label{c}
Pr\left(X_1 \geq 2 \mu_0\right) > 1- e^{-\mu_0/3}
\end{equation}
That is, the probability of starting with at least $2 \mu_0$ values of 1 is more than $1- e^{-\mu_0/3}$.
\qed

\subsection*{Putting it all together}
First, in Lemma~\ref{lemma:randomWalk},
we computed the probability that the final decision value is the majority value $1$,
\emph{conditioned} on starting with at least $2 \mu_0$ values of $1$.
Then, in Lemma~\ref{lemma:usingChernoff}, we computed the probability of starting with at least $2 \mu_0$ values of 1.
Multiplying these two probabilities gives us the probability that the final decision value is 1. Thus,
assuming $W_1 > W_0$, the probability that the final decision value is 1 is more than
$$
\left(1-(W_0/W_1)^{4 \mu_0}\right) \times \left(1- e^{- \mu_0 /3}\right)
$$
%
This completes the proof of Theorem~\ref{thm:majority}. \qed

\section{Analysis: Computing the expected number of steps}
\label{subsec:analysis:write-erase}
Recall that it is assumed that $W_1 = E_0$ and $W_0 = E_1$.
Next, we compute the expected number of big steps (i.e., time units)
needed to until reaching agreement, when executing the epigenetic consensus algorithm.
(Time complexity is defined in Section~\ref{sec:model}.)
\begin{theorem}
\label{thm:expectedRunTime}
Let $T$ be the number of steps needed to reach agreement (on either 0 or 1).
Then, assuming $W_1 > W_0$,
$$
E[T] \leq \frac{2(W_0 + W_1)^4 N^2}{W_1^4 - W_0^4} .
$$
\end{theorem}

\begin{corollary}
Assuming $W_1 / W_0 \geq 3$, $E[T] \leq 6.4 N^2$ .
\end{corollary}

\setcounter{equation}{0}
\proof
Recall that an \emph{update-steps} is an attempt to resolve a collision,
in which a value is erased and then a value is written in the same location.
An update-step may result a change in the number of 1's (and hence also of 0's).
There are three such types of update-steps which we have named \emph{win}, \emph{lose} and \emph{tie}.
A win step, which can happen with probability $P$,
is when a collision is changed into 11.
A lose step, which can happen with probability $Q$,
is when a collision is changed into 00.
A tie step, which can happen with probability $R$,
is when a collision is not changed.
The values of $P$, $Q$ and $R$ are as in the previous section.

We first focus on computing the expected number of \emph{update-steps} of the epigenetic consensus algorithm.
This question is identical to the question about the expected number of steps in the gambler's ruin problem from Section~\ref{sec:AnalysisPreliminaries}. As explained in the previous section,
going broke is analog to reaching agreement on 0, where going home winner with a fortune of
$N$ is analog to reaching agreement on 1.
Winning one unit with probability $p$ is analog to a win step with probability $P$,
losing one unit with probability $q$ is analog to a lose step with probability $Q$, and
not losing or winning with probability $r$ is analog to a tie step with probability $R$.
Finally, the gambler starting with $i$ units is analog to assuming that the initial number of 1's is $i$.

Let $U_i$ be the number of \emph{update-steps} until consensus is reached,
starting with initially $i$ values of $1$, $0\leq i\leq N$.
By Lemma~\ref{lemma:Expected Playing Time},
\begin{equation}\label{AAA}
E[U_i] = \left( \left( \frac{N}{P-Q}\right) \left[ \frac{1-(Q/P)^i}{1-(Q/P)^N} \right] - \frac{i}{P-Q} \right) \left( \frac{1}{P+Q} \right)
\end{equation}
The assumption that $W_1 >W_0$ implies that $P>Q$, and thus we can simplify (\ref{AAA}) as follows,
\begin{equation}\label{BBB}
E[U_i] \leq \left( \frac{N}{P-Q} - \frac{i}{P-Q} \right) \left( \frac{1}{P+Q} \right)
      = \frac{N-i}{P^2 - Q^2}
\end{equation}
Since $P= (W_1/(W_0 + W_1))^2$ and $Q= (W_0/(W_0 + W_1))^2$, we get that
\begin{equation}\label{CCC}
E[U_i] \leq \frac{(W_0 + W_1)^4 (N-i)}{W_1^4 - W_0^4}
\end{equation}
Let $U$ be the number of \emph{update-steps} until consensus is reached.
Then,
\begin{equation}\label{DDD}
E[U] \leq \max_{i\in\{1,...,N-1\}} {E[U_i]} \leq \frac{(W_0 + W_1)^4 (N-1)}{W_1^4 - W_0^4}
\end{equation}
So far we have computed, $E[U]$, the expected number of \emph{update-steps} of the epigenetic consensus algorithm.
Next, we compute the expected number of (big) steps in general.
For update-steps to take place, we need the erasers and writers to
arrive at the collision locations.
So, for a single update-step, in the worst case, we may need to wait
for $N-1$ steps until the eraser arrives and an addition $N-1$ steps
until the writer arrives. Thus, for each update-steps, in the worst case, we should count
$2N-2$ additional steps. That is, counting a total of $2N-1$ steps for each update-step.
In addition, in the worst case, we should add $N$ (big) steps for the first writes into the $N$ memory location.
Thus, from (\ref{DDD}) and as explained above, it follows that
\begin{equation}\label{EEE}
E[T] \leq \frac{(W_0 + W_1)^4 (N-1)}{W_1^4 - W_0^4} \times2(N-1) +N \leq \frac{2(W_0 + W_1)^4 N^2}{W_1^4 - W_0^4}
\end{equation}
This completes the proof of Theorem~\ref{thm:expectedRunTime}.
\qed

\begin{figure}[tb]
\centering
\includegraphics[scale=.41]{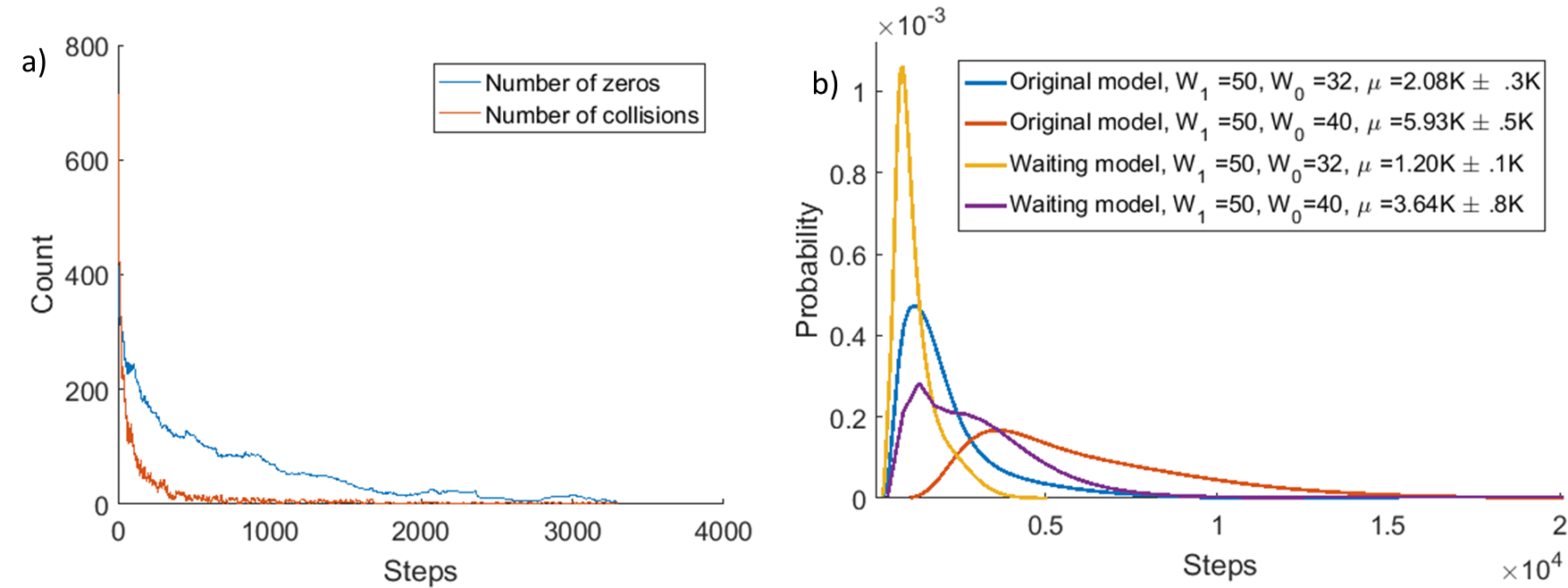}
\caption{a) Number of zeros and collisions vs steps in the algorithm. While the initial number of collisions is a linear function of the number of 0's, we observe that towards the end of the algorithm there are very few collisions while the number of 0's remains relatively high. b) Comparison between the proposed original model and a revised model that allows writers to attach themselves with the erasers and the erasers wait until a collision is resolved. Here we can see that the waiting version is faster at both competition level compared to the original model.}\label{zer_coll}
\end{figure}

\textbf{Remark: }
The analysis leads to a runtime  of $O(N^2)$ steps. However, it  assumes a single collision being resolved each time. In practice, we have multiple collisions which can resolved in parallel. Initially, the number of collisions is a linear function of the number of 0's. If the number of collisions remains a linear function of the number of 0's then in $O(N)$ steps we would resolve $O(N)$ collisions, which would lead to at most $O(N \lg N)$ runtime. Unfortunately, as Fig. ~\ref{zer_coll}a) shows based on simulations, this is not the case. The epigenetic consensus algorithm leads to a rapid decrease in the number of collisions while not decreasing the number of 0's at the same rate. This means that long stretches of 0's (and 1's) form, leading to a small number of collisions while still having a large number of 0's.
One way to overcome this is to change the algorithm to better mimic what biology does. Specifically, in biology we observe that erasers and writers interact during the establishment of a new state \cite{torres2015functional}. We hypothesize that an algorithm that utilizes these ideas can indeed lead to a faster runtime as the simulation analysis below shows.

\begin{figure}[tb]
\centering
\includegraphics[scale=.41]{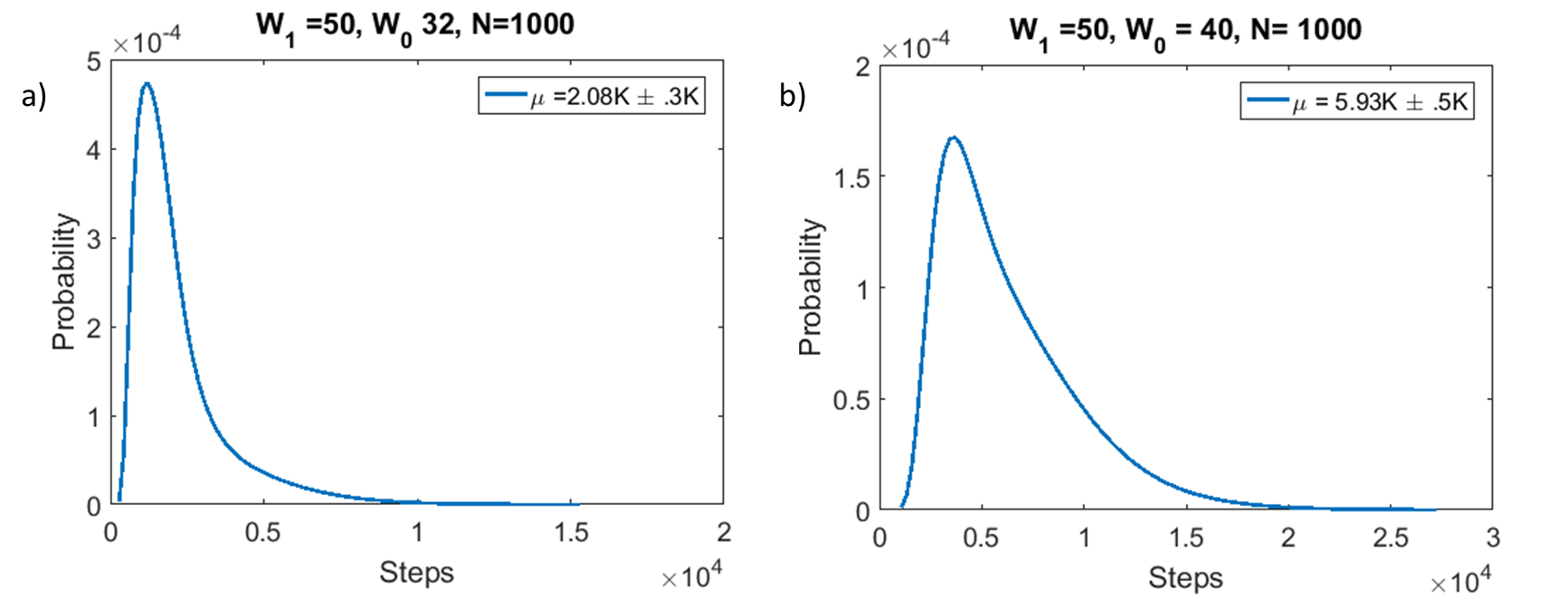}
\caption{Distribution of number of steps to reach consensus. Plots summarize 300 random runs of the algorithm. a) low and b)high level of competitions between 1-writers and 0-writers. $\mu$ denotes the average time to reach consensus.}\label{sim1}
\end{figure}

\section{Simulations and analysis of real biological data}
\label{sec:Simulations}
We performed simulations of our proposed algorithm at two different levels of competitions between the 1-writers and 0-writers. Fig.~\ref{sim1} shows the probability distribution of the number of steps to reach consensus. To simulate high level of competition we used $W_1=50, W_0=40, p= 0.56$. To simulate low level of competition we used $W_1=50, W_0=32, p=0.60$. For both cases, $N=1000$. As expected, at low level of competition consensus is achieved much faster. The average number of steps taken to reach consensus is $2.08K$ compared to $5.93K$ for the high level of competition.

Fig.~\ref{zer_coll}(a) shows the change in the number of zeros and collisions as the algorithm progresses towards consensus. As can be seen from the figure, initially the number of collisions is linear in the number of 0's. However, the algorithm proposed leads to a quick drop in the number of collisions while the number of 0's remains fairly high which means that collisions cannot be resolved in parallel. We also simulated an alternative, which allows 1-writers to attach to 0-erasers (forming a writer-eraser complex though not guaranteeing atomicity). As before, erasers continue to scan the DNA until they reach a collision. However, in the revised version erasers wait at the collision site to see what value was written and if the new value leads to another collision they attempt to erase again until  the collision is resolved. As mentioned above, if the ratio of erasers from the two types is not 1, it would always converge to a consensus. While we are unable to prove a better worst case runtime for such a method, simulations indicate that it leads to much faster convergence when compared to the epigenetic consensus algorithm (Fig.~\ref{zer_coll}b). Coupled with recent biological observations \cite{panneerdoss2018cross} these results indicate that this method is likely much faster than our current proposed algorithm while still not requiring any memory for the processors.

\begin{figure}[!tb]
\centering
\includegraphics[scale=.4]{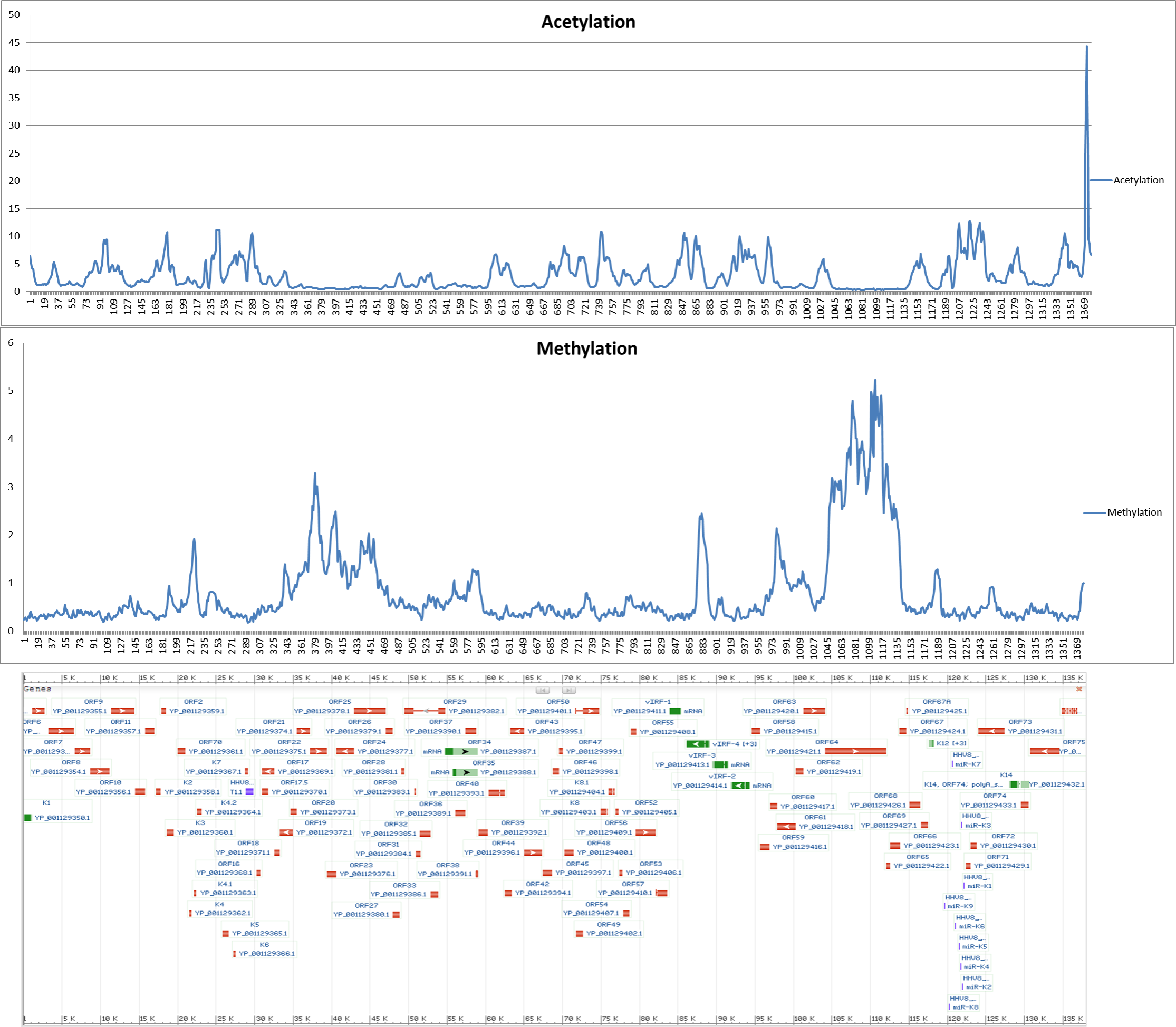}
\caption{Different histone modifiers competing for the same residue, Histone 3 Lysine 9 (H3K9). Latent Kaposi Sarcoma-Associated Herpesvirus Genomes (Resolution 250 bp). Two types of histone modifiers are competing to put acetylation and methylation marks on the residue. We can see stretches of regions with only methylation or acetylation marks, showing regional consensus.}\label{tracks}
\end{figure}

We have also looked at recent epigenetic data to see if the assumptions we made about increasing stretches using collisions rather than randomly changing location are observed in real data. Fig. ~\ref{tracks}
presents results from a recent study by Gunther \textit{et al} \cite{gunther2010epigenetic} in which two different types of marks, acetylation and methylation are competing for the same residue H3K9 (and so can be thought of as 0 and 1). As can be seen, and in agreement with our local consensus formulation, there is regional consensus of these marks. The methylation mark is strong in the region that lies between 30K bps to 60K bps and again from 105k bps to 120K bps. The acetylation marks are almost non existent in these regions, demonstrating regional consensus of methylation marks in the DNA and the likely impact of collisions on the establishement of such regions.

\begin{figure}
\centering
\includegraphics[scale=.31]{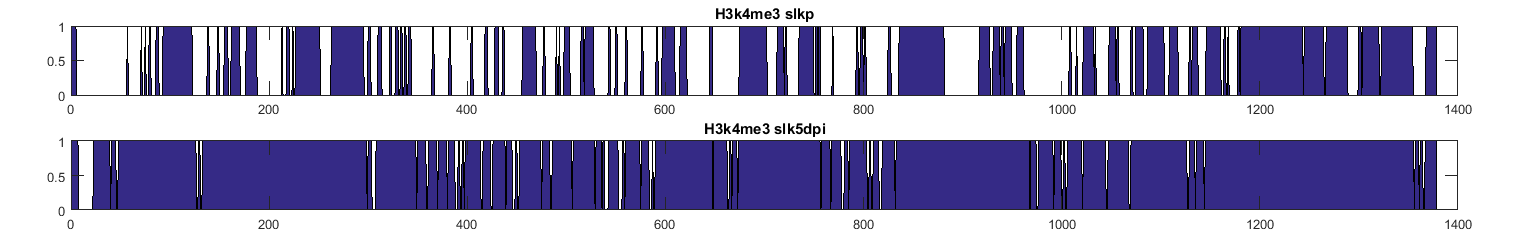}
\caption{Increase in histone modification intensity from the initial stage of infection (slkp) to 5 days post infection (5dpi). We can see that the later time point has higher overall consensus of methylation histone mark.}\label{consensus}
\end{figure}

These consensus states are dynamically regulated in response to stress or for establishing a new state during development. Fig.~\ref{consensus} shows results from a temporal study by Gunther \textit{et al} \cite{gunther2010epigenetic} in which cells are gradually moving towards consensus of tri-methylation marks of Histone 3 Lysine 4 (H3K4) residue as a response to Kaposi Sarcoma infection. The blue lines indicate presence of the H3K4me3 mark. The figure presents two time points, the first is at onset of infection (when the virus is applied) while the second one is from a sample 5 days post infection. As can be seen, after 5 days we observe a higher level of consensus than at the onset of infection indicating that collisions continue to be resolved until full consensus is reached. Most likely, in this case, the reason for the increase is the activation of  additional modifiers at the later time points which lead to changes in the ratio of 1-writers and 0-writers and erasers. We note that the studies performed to-date are looking at a collection at cells at once and so only report average data. New technologies are enabling us, for the first time, to observe these events at a single cell resolution \cite{cheung2018single} and we expect that these results will further help us infer the specific algorithms utilized by cell to reach consensus.

\section{Discussion}
\label{sec:Discussion}
To date, the study of ``algorithms in nature'' at the molecular and cellular levels, i.e., how collections
of molecules and cells process information and solve computational problems,
was discussed mainly in the context of networks and message passing \cite{navlakha2015distributed}.
This paper attempts
to study biologically inspired distributed computing algorithms
in the context of molecular shared memory systems.

We have focused on the process of genome-wide epigenetic modifications in which cells utilize DNA as a shared memory object to establish a new state. We formulated the new epigenetic consensus problem that these modifiers need to solve and presented algorithms and their  expected run time. We have also discussed and simulated improved methods for solving the problem which rely on additional recent biological insights. By analyzing real biological data we show that the decisions made in the algorithms we presented, to focus on collisions, indeed reflect experimental results for the establishment of new cell states using epigenetics.

Robustness is a desired property of cellular and molecular systems
That is, they should be able to recover and restore their original state after a disturbance
(a transit failure) without any outside intervention.
We note that our epigenetic consensus algorithm is robust.
It can easily tolerate a limited number of arbitrary memory location (value) changes and processor failures.

Our consensus algorithm is ``one-shot,'' and we would also like to cover the ``long-lived'' version
in which we may switch again possibly many times (i.e., repeated consensus). Also, we assume
that all the shared memory locations are initially empty and it would be nice to be able to remove this assumption,
especially when a need to establish a new state arises.
A tiny change to our algorithm achieve the above desired properties.
This is done by assuming that a $v$-eraser ``once in a while'' (i.e., at random)
unconditionally erase $v$, even when there is no collision. In such case, if the ratio between
the number of 0-erasers and 1-erasers changes significantly, the decision value will change as well.
Thus, with this tiny change (which is biologically justified) the algorithm is self-stabilizing --
agreement is reached starting from any configuration (i.e., from any assignment of values to the memory locations).

There are several, possibly faster, variants of our epigenetic consensus algorithm that
are theoretically interesting, but require making assumptions that are not
acceptable from a biological standpoint. For example, we may assume that
each writer is in one of two states: active or inactive. Initially, all writers are active.
In an active state, a writer behaves as before (scans and writes in empty locations).
In an inactive state, a writer scans the array but never writes.
Let $k_1$ and $k_2$ be small positive integers.
An active  (resp.\ inactive) $v$-writer become inactive (resp.\ active)
if the value in all the last $k_1$ (resp.\ $k_2$) locations it has visited is $1-v$ (resp.\ $v$).
Such a solution requires each writer to have a few additional bits of local memory, which is
biologically unrealistic.

Finally, we point out that the simplicity of our proposed algorithm
and its analysis does not mean they were simple to obtain.
The following quote is from a letter written by
the French mathematician and philosopher Blaise Pascal in 1657
``I have made this letter longer than usual, only because I have not had the time to make it shorter.''
We had enough time to obtain a simple algorithm and a simple analysis.

\bibliographystyle{plain}
\bibliography{BibBiology}

\end{document}